\documentclass[sigconf,screen]{acmart}

\usepackage{enumitem}
\usepackage{fontawesome}

\AtBeginDocument{%
  }

\copyrightyear{2026}
\acmYear{2026}
\setcopyright{cc}
\setcctype{by}
\acmConference[MSR '26]{23rd International Conference on Mining Software Repositories}{April 13--14, 2026}{Rio de Janeiro, Brazil}
\acmBooktitle{23rd International Conference on Mining Software Repositories (MSR '26), April 13--14, 2026, Rio de Janeiro, Brazil}
\acmPrice{}
\acmDOI{10.1145/3793302.3793603}
\acmISBN{979-8-4007-2474-9/2026/04}

\usepackage{subcaption}

\definecolor{framebg}{gray}{0.95}
\newsavebox{\framedboxsave} 

\newenvironment{framed}{
  \par\smallskip
  \begin{lrbox}{\framedboxsave}
  \begin{minipage}{\linewidth}
}{
  \end{minipage}
  \end{lrbox}
  \noindent\colorbox{framebg}{\usebox{\framedboxsave}}
  \par\smallskip
}

\usepackage{fancybox}
\newenvironment{keytakeaway}{
  \par\smallskip
  \noindent
  \begin{minipage}{\linewidth}
  \hrule\vspace{4pt}
  \small
}{
  \vspace{4pt}\hrule
  \end{minipage}
  \par\smallskip
}

\begin{document}


\title{How AI Coding Agents Modify Code: A Large-Scale Study of GitHub Pull Requests}
\author{Daniel Ogenrwot}
\orcid{0000-0002-0133-8164}
\affiliation{%
  \institution{University of Nevada Las Vegas}
  \city{Las Vegas}
  \state{Nevada}
  \country{USA}
}
\email{ogenrwot@unlv.nevada.edu}

\author{John Businge}
\orcid{0000-0003-3206-7085}
\affiliation{%
  \institution{University of Nevada Las Vegas}
  \city{Las Vegas}
  \state{Nevada}
  \country{USA}
}
\email{john.businge@unlv.edu}

\renewcommand{\shortauthors}{Daniel and John}

\begin{abstract}
AI coding agents are increasingly acting as autonomous contributors by generating and submitting pull requests (PRs). However, we lack empirical evidence on how these agent-generated PRs differ from human contributions, particularly in how they modify code and describe their changes. Understanding these differences is essential for assessing their reliability and impact on development workflows. Using the MSR 2026 Mining Challenge version of the AIDev dataset, we analyze 24{,}014 merged Agentic PRs (440{,}295 commits) and 5{,}081 merged Human PRs (23{,}242 commits). We examine additions, deletions, commits, and files touched, and evaluate the consistency between PR descriptions and their diffs using lexical and semantic similarity. Agentic PRs differ substantially from Human PRs in commit count (Cliff's $\delta = 0.5429$) and show moderate differences in files touched and deleted lines. They also exhibit slightly higher description-to-diff similarity across all measures. These findings provide a large-scale empirical characterization of how AI coding agents contribute to open source development.

\end{abstract}






\maketitle

\section{Introduction}
Artificial intelligence (AI) coding agents are rapidly transforming software development. Tools such as GitHub Copilot~\cite{githubCopilot}, OpenAI Codex~\cite{openaiCodexWeb}, Claude Code~\cite{claudeAI2025}, Cursor~\cite{cursor2025}, and Devin~\cite{devinAIWeb} can now autonomously generate code, fix bugs, and submit pull requests (PRs). These systems are evolving from assistive tools into active collaborators, a transition towards Software Engineering 3.0~\cite{hassan2025agenticsoftwareengineeringfoundational,hassan2024ainativesoftwareengineeringse}. Prior work has examined how developers interact with AI-generated suggestions and how such tools affect productivity and code quality~\cite{vaithilingam2022expectation,li2025aidev,vaithilingam2023copilot,ogenrwot2025patchtrackcomprehensiveanalysischatgpts}, highlighting both opportunities and challenges for human--AI collaboration.

Despite this growing adoption, we still lack empirical evidence about how AI-generated PRs differ from human-authored ones in real-world repositories. In particular, little is known about how coding agents modify source code or how accurately their PR descriptions reflect the associated edits. These gaps limit our ability to assess their reliability, maintainability impact, and communicative clarity during code review. While existing research focuses primarily on usage patterns and developer interactions~\cite{barke2023grounded,vaithilingam2023copilot,10.1145/3613904.3642349}, large-scale analyses of autonomous AI-authored PRs remain scarce.

The MSR 2026 Mining Challenge dataset~\cite{li2025aidev} enables this study by providing thousands of merged Agentic and Human PRs with commit histories and diff information. We therefore examine two research questions: \faQuestionCircle \hspace{0.02in} \textbf{RQ1:} How do the structural characteristics of Agentic PRs, such as additions, deletions, files touched, and commit count, compare to Human-authored PRs? \faQuestionCircle \hspace{0.02in} \textbf{RQ2:} How does the alignment between PR descriptions and code edits differ between Agentic and Human PRs?

These questions help reveal how AI agents behave as contributors and how effectively they communicate intent through natural language, both of which are critical for trust and collaboration in software engineering workflows.

\noindent \textbf{Contributions.}
\begin{itemize}[leftmargin=*]
    \item We conduct a large-scale empirical comparison of the structural code changes of Agentic and Human PRs using the MSR 2026 Mining Challenge dataset.
    \item We assess description-to-diff alignment to evaluate how well PR descriptions capture the underlying code modifications.
    \item We release a replication package containing curated data and analysis scripts to support reproducibility~\cite{replication-package}.
\end{itemize}
\section{Method}
\begin{figure*}[ht]
\vspace{-5pt}
    \centering
    \includegraphics[width=0.90\linewidth]{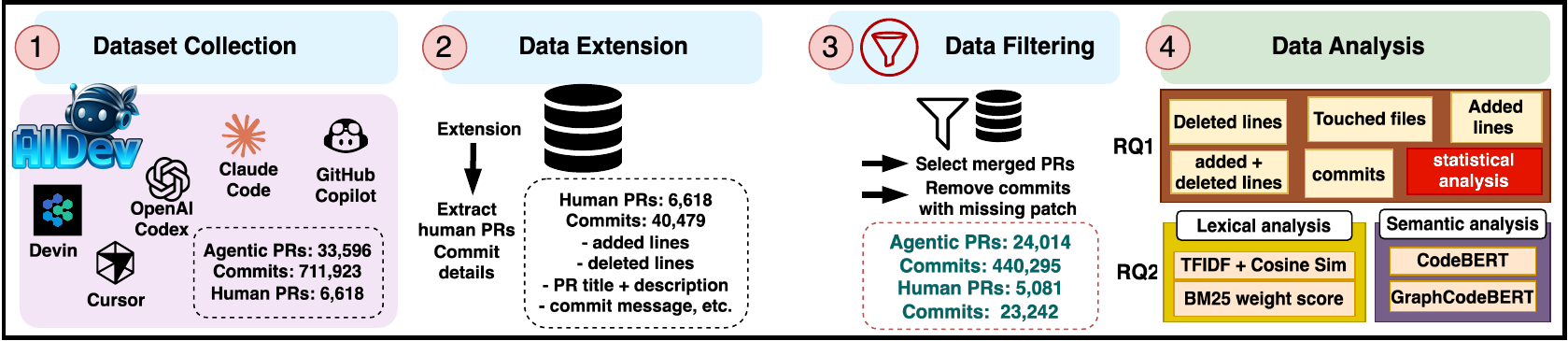}
    \captionsetup{skip=0.5pt}
    \caption{Four-step workflow: dataset collection, commit-data extension, PR filtering, and structural and similarity analysis.}
    \Description{Four-step workflow: dataset collection, commit-data extension, PR filtering, and structural and similarity analysis.}
    \label{fig:method}
    \vspace{-10pt}
\end{figure*}

Figure~\ref{fig:method} summarizes our four-step process: dataset collection, extension of Human PRs with commit-level data, filtering of merged PRs with valid patches, and computation of structural and similarity metrics for analysis.

\noindent \textbf{Step 1: Dataset Collection.}
We use the AIDev dataset~\cite{li2025aidev} provided for the MSR 2026 Mining Challenge (retrieved November 1, 2025). The dataset contains 932{,}791 Agentic PRs and 6{,}618 Human PRs across 116{,}211 repositories, along with a curated subset enriched with comments, reviews, and issue links. We use the Agentic PR set for structural analysis and reconstruct commit-level details for the Human PRs to enable consistent comparison.

\noindent \textbf{Step 2: Extracting Human-PRs Commit Details.}
Unlike Agentic PRs, the Human PRs in the AIDev dataset lack commit-level information. To enable consistent comparison, we retrieved commit metadata and file-level patches for all Human PRs using the GitHub REST API~\cite{GitHubRestAPI2025}. For each PR, we collected its commits and their associated details, including modified files, additions, deletions, and the unified diff when available. This reconstruction yields a commit-level structure matching that of the Agentic PRs. The number of files touched is computed as the set of unique file paths modified across all commits, ensuring repeated edits to the same file are counted once. File renames reported by GitHub are treated as single-file modifications.

\noindent \textbf{Step 3: PR Filtering.}
We focus on merged PRs, which provide complete commit histories. For Agentic PRs, we removed entries lacking patch text, required for computing structural and similarity metrics. A commit was considered to have a valid patch if the GitHub API returned a non-empty unified diff for all modified files. For Human PRs, we excluded cases with missing repositories, unavailable commit data, or incomplete patches retrieved from the GitHub API. After filtering, the analysis set contains 24{,}014 Agentic PRs (440{,}295 commits with valid patches) and 5{,}081 Human PRs (23{,}242 commits with valid patches). These datasets form the basis of all analyses in RQ1 and RQ2. Agentic PRs are also grouped by agent type for stratified analyses in the results, following established large-scale mining practices~\cite{businge:emse:2022,businge:saner:2022,Businge:benevol:2020,businge:2018icsme}.

\noindent \textbf{Step 4: Data Analysis.}
For RQ1, we computed code-change metrics for each PR, including the number of commits, total lines added and deleted, and files touched. 
Commit-level additions and deletions were summed across all commits in a PR, which avoids double-counting because GitHub reports per-file statistics. 
We summarized the distributions of these metrics for Agentic and Human PRs. Normality assumptions were violated for all metrics, motivating the use of the Mann--Whitney $U$ test~\cite{Mann-Whitney, mann-whitney-2} to assess differences between the two groups. Effect sizes were quantified with Cliff’s delta ($\delta$) using standard interpretation thresholds~\cite{long2003ordinal,romano2006exploring}.

For RQ2, we evaluate description--diff alignment along two dimensions. The first is \textit{lexical similarity}, which measures surface-level overlap between the PR description and the code diff. We compute Term Frequency--Inverse Document Frequency (TF--IDF) cosine similarity and Okapi BM25, two standard information retrieval baselines widely used in software engineering tasks, particularly in bug report similarity and localization~\cite{SALTON1988513, robertson-2009-probabilistic, Yang2016-word-embedding, Samir2025ImprovedIR}. PR titles and descriptions are concatenated and normalized. For all lexical and semantic analyses, code diffs are lowercased and stripped of diff metadata (file headers \texttt{+++}/\texttt{---}, hunk headers \texttt{@@}, and leading \texttt{+}/\texttt{-}), retaining only added and deleted lines with normalized whitespace. 

Both PR description and code diff are then tokenized using a simple word-based tokenizer~\cite{10.5555/1717171}. Let $d$ denote the normalized description and $c$ the cleaned diff; TF--IDF cosine similarity is computed between vectors $\mathbf{v}_d$ and $\mathbf{v}_c$, treating $c$ as the query to capture whether tokens introduced or modified by the patch are reflected in the description.

The second dimension is \textit{semantic similarity}, which evaluates whether the meaning expressed in the PR description aligns with the underlying code edits. Embeddings are computed for both texts using CodeBERT~\cite{feng2020codebertpretrainedmodelprogramming} and GraphCodeBERT~\cite{guo2021graphcodebertpretrainingcoderepresentations}, and similarity is measured using cosine similarity, a standard choice for comparing embedding vectors in semantic similarity settings~\cite{reimers2019sentencebertsentenceembeddingsusing}. We include both CodeBERT and GraphCodeBERT to capture semantic similarity under distinct model architectures: a masked language model and a data-flow--aware encoder. RQ2 is evaluated on the subset of PRs that contain complete description text and valid diff data after filtering in Step~3.

TF--IDF and embedding cosine scores are bounded (typically in $[0,1]$), with larger values indicating stronger alignment. Okapi BM25, in contrast, is an unbounded relevance score whose magnitude depends on token frequencies and document length; long and token-heavy diffs can therefore produce very large positive or negative values. Negative Okapi BM25 values arise when very common tokens receive low inverse document frequency weights, which can reduce the summed relevance score for long diffs. As a result, Okapi BM25 should be interpreted only as a relative lexical signal rather than a calibrated similarity measure~\cite{robertson-2009-probabilistic,10.1007/978-3-642-28997-2_21}. Because absolute cutoffs are task-dependent, interpretation relies on the score distributions (e.g., medians and quartiles).
\vspace{-5pt}
\section{Results}
\begin{framed}
\faQuestionCircle \hspace{0.02in} \textbf{RQ 1: How do the structural characteristics of Agentic PRs, such as additions, deletions, files touched, and commit count, compare to Human-authored PRs?}
\end{framed}

\begin{figure}[ht]
\vspace{-8pt}
    \centering
    \includegraphics[width=0.85\linewidth]{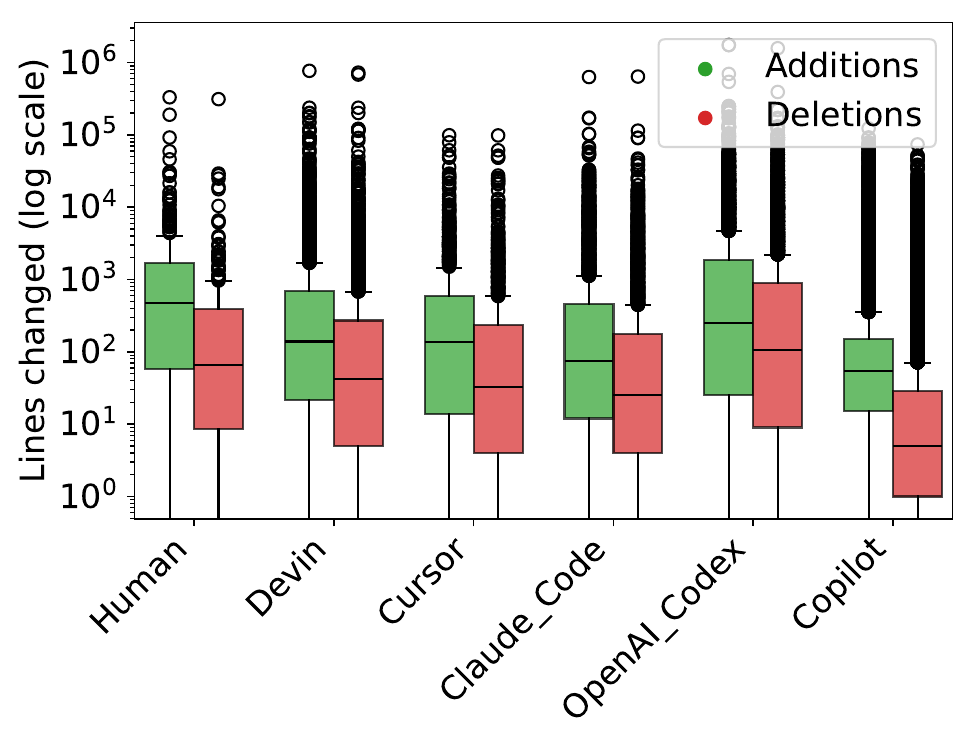}
    \captionsetup{skip=0pt}
    \caption{Distribution of LOC added and deleted in agentic vs.\ human PRs.}
    \Description{Distribution of LOC added and deleted in agentic vs.\ human PRs.}
    \label{fig:boxplot-lines-changed}
    \vspace{-6pt}
\end{figure}

\noindent Figure~\ref{fig:boxplot-lines-changed} shows that Human PRs exhibit the largest and most variable code changes, with substantially higher medians and long upper tails for both additions and deletions. Agentic PRs tend to introduce smaller and more localized edits. Within the Agentic group, Claude Code and OpenAI Codex display somewhat greater variability in their additions, whereas Devin, Cursor, and especially Copilot produce more consistently small and localized changes.

\begin{keytakeaway}
\small
\faHandORight \hspace{0.02in} \textit{\textbf{Key takeaway:} Agentic PRs tend to introduce smaller and more localized changes than Human PRs, though some agents (e.g. Claude Code and OpenAI Codex) show broader variability than others.}
\end{keytakeaway}

\begin{figure}[ht]
    \centering
    \includegraphics[width=0.8\linewidth]{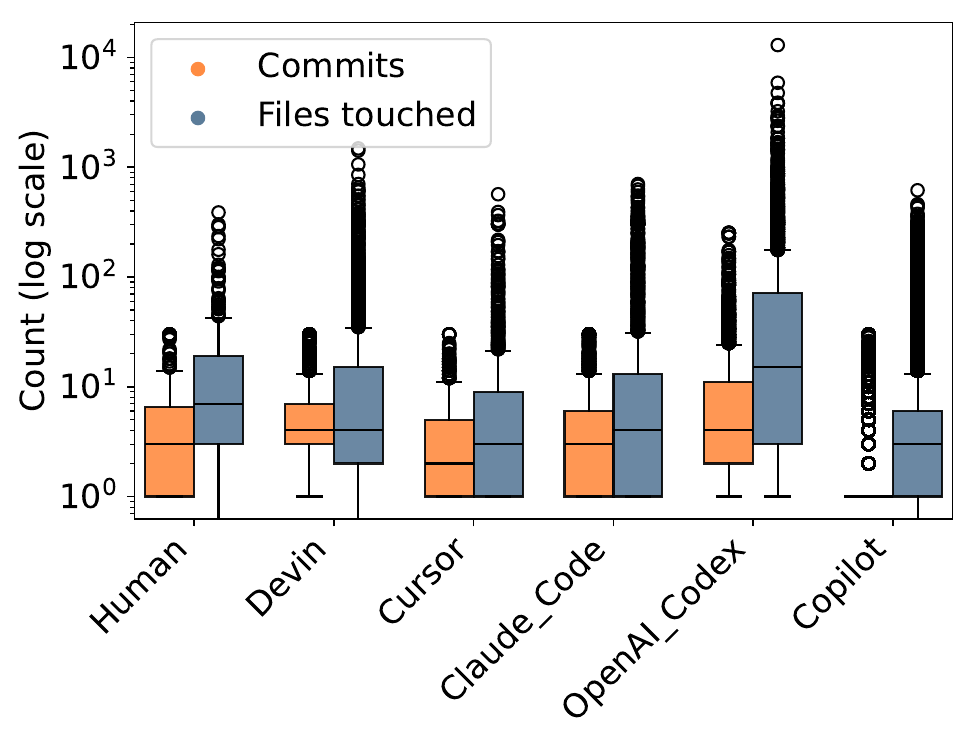}
    \captionsetup{skip=0pt}
    \caption{Distribution of commits and files touched across agents and human PRs.}
    \Description{Distribution of commits and files touched across agents and human PRs.}
    \label{fig:boxplot-commits-filetouched}
    \vspace{-14pt}
\end{figure}

Figure~\ref{fig:boxplot-commits-filetouched} shows that Human PRs touch more files and often involve more commits than Agentic PRs, indicating broader and more distributed modifications. Claude Code and OpenAI Codex again show a wider spread than other agents, while Devin, Cursor, and Copilot exhibit narrow distributions that reflect highly localized edits. Human contributions consistently span more files and commit steps, matching the large effect size observed for commit count. 

\begin{keytakeaway}
\small
\faHandORight \hspace{0.02in} \textit{\textbf{Key takeaway:} Agentic PRs are not uniform; some tools generate highly compact, single-focus contributions, while others produce broader edits that overlap more closely with the lower range of Human PRs.}
\end{keytakeaway}

\begin{table}[ht]
\vspace{-0.3cm}
\small
\centering
\caption{Cliff's $\delta$ Effect Sizes - agentic vs.\ human PRs}
\vspace{-12pt}
\begin{tabular}{lcc}
\toprule
\textbf{Metric} \hspace{4cm} & \textbf{Cliff's $\delta$} & \textbf{Effect Size} \\
\midrule
Commits & 0.5429 & Large \\
Files Touched & 0.4487 & Medium \\
Additions     & 0.2836 & Small \\
Deletions     & 0.4462 & Medium \\
Line Changes (add $+$ delete) & 0.3158 & Small \\
\bottomrule
\end{tabular}
\label{tab:cliffs-delta}
\vspace{-8pt}
\end{table}

Table~\ref{tab:cliffs-delta} shows that the largest practical difference between Agentic and Human PRs is the number of commits ($\delta = 0.54$, large effect), indicating distinct contribution patterns in how changes are structured. Medium effects for files touched and deletions suggest that Human PRs modify codebases more broadly and remove more code. In contrast, additions and total line changes show only small effects, suggesting that the overall quantity of code introduced is closer between the two groups than the structure of the edits. Overall, the result of the Mann–Whitney $U$ test shows that the differences are statistically significant ($p \le 0.001$).

\begin{keytakeaway}
\small
\faHandORight \hspace{0.02in} \textit{\textbf{Key takeaway:} The most distinctive characteristics of Agentic PRs are not simply how many lines they change, but how they organize and distribute changes across commits and files.}
\end{keytakeaway}

\begin{framed}
\faQuestionCircle \hspace{0.02in} \textbf{RQ 2: How does the alignment between PR descriptions and code edits differ between Agentic and Human PRs?}
\end{framed}

\begin{figure}[ht]
\vspace{-8pt}
    \centering
    \includegraphics[width=0.8\linewidth]{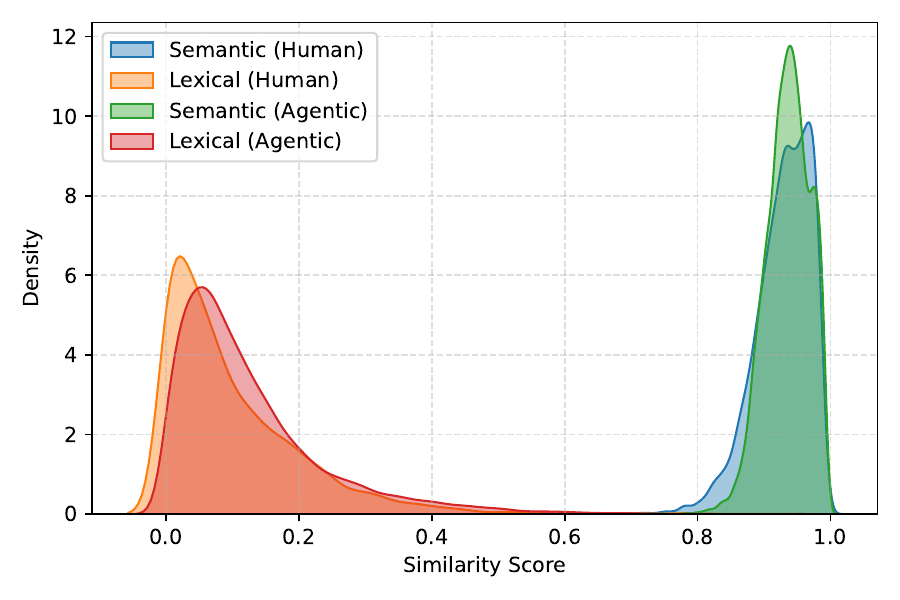}
    \captionsetup{skip=0pt}
    \caption{Kernel density estimates comparing lexical and semantic similarity distributions.}
    \Description{Kernel density estimates comparing lexical and semantic similarity distributions}
    \label{fig:RQ2_agg_kde}
    \vspace{-10pt}
\end{figure}

\begin{figure*}[ht]
    \centering
    \includegraphics[width=0.9\linewidth]{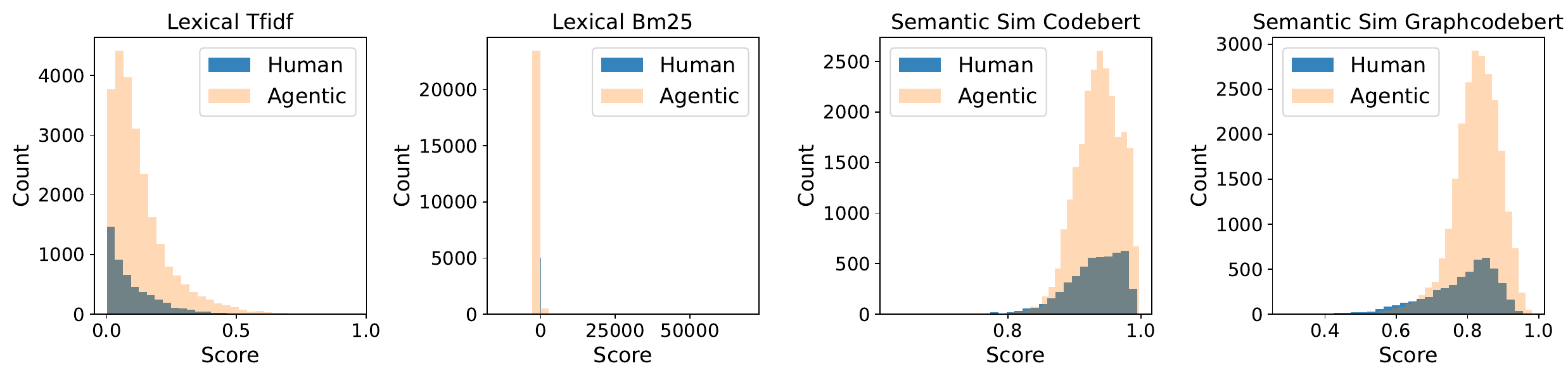}
    \captionsetup{skip=0pt}
    \caption{Distribution of similarity scores across lexical and semantic metrics for both Agentic and Human PRs.}
    \Description{Distribution of similarity scores across lexical and semantic metrics for both Agentic and Human PRs.}
    \label{fig:RQ2_agg_histograms}
    \vspace{-5pt}
\end{figure*}

\begin{table*}[ht]
\small
\centering
\caption{Descriptive statistics for lexical and semantic similarity metrics, grouped by PR Type.}
\vspace{-12pt}
\label{tab:combined-side-by-side}
\begin{tabular}{l *{2}{cccc}}
\toprule
\textbf{Statistic} & \multicolumn{4}{c}{\textbf{Agentic PRs}} & \multicolumn{4}{c}{\textbf{Human PRs}} \\
\cmidrule(lr){2-5} \cmidrule(lr){6-9}
& \multicolumn{2}{c}{\textbf{Lexical}} & \multicolumn{2}{c}{\textbf{Semantic}}
& \multicolumn{2}{c}{\textbf{Lexical}} & \multicolumn{2}{c}{\textbf{Semantic}} \\
\cmidrule(lr){2-3} \cmidrule(lr){4-5} \cmidrule(lr){6-7} \cmidrule(lr){8-9}
& TF--IDF & Okapi BM25 & CodeBERT & GraphCodeBERT
& TF--IDF & Okapi BM25 & CodeBERT & GraphCodeBERT \\
\midrule
Mean
& \colorbox{lightgray}{\textbf{0.1245}} & \colorbox{lightgray}{\textbf{2.8455}} & \colorbox{lightgray}{\textbf{0.9356}} & \colorbox{lightgray}{\textbf{0.8254}}
&  0.1007 & 0.1739 & 0.9285 & 0.7815 \\
Std.
& 0.1104 & 466.2787 & 0.0331 & 0.0654
& 0.1054 & 23.1437 & 0.0423 & 0.1016 \\
Min
& 0.0000 & -13776.6311 & 0.7243 & 0.4183
& 0.0000 & -1271.3196 & 0.6259 & 0.2875 \\
25\%
& 0.0479 & -0.3836 & 0.9141 & 0.7889
& 0.02413 & -0.2380 & 0.9042 & 0.7254 \\
50\%
& \colorbox{lightgray}{\textbf{0.0937}} & \colorbox{lightgray}{\textbf{1.0169}} & \colorbox{lightgray}{\textbf{0.9375}} & \colorbox{lightgray}{\textbf{0.8302}}
& 0.0680 & 0.0000 & 0.9347 & 0.8067 \\
75\%
& 0.1655 & 1.4011 & 0.9605 & 0.8708
& 0.1465 & 1.1236 & 0.9617 & 0.8563 \\
Max
& 0.9535 & 68947.5283 & 0.9974 & 0.9807
& 0.9266 & 585.4945 & 0.9939 & 0.9788 \\
\bottomrule
\end{tabular}
\vspace{-10pt}
\end{table*}

We first examine the distribution of lexical and semantic similarity scores for Human and Agentic PRs. Figure~\ref{fig:RQ2_agg_kde} shows a clear separation between lexical and semantic similarity. Lexical scores for both Agentic and Human PRs cluster near zero, indicating very limited surface-level vocabulary overlap between descriptions and diffs. In contrast, semantic similarity scores form tight peaks between 0.9 and 1.0, suggesting that both description types capture the underlying meaning of their patches even when wording differs.

Figure~\ref{fig:RQ2_agg_histograms} presents distributional comparisons across all four metrics. Agentic PRs show a slight right shift relative to Human PRs across both lexical and semantic measures, although differences are small and both groups exhibit high semantic alignment. Lexical measures remain heavily skewed, while semantic models produce consistently high similarity scores, with Agentic PRs showing tighter clustering. Okapi BM25 exhibits very high variance and large positive or negative values, particularly for Agentic PRs, reflecting its unbounded nature and sensitivity to tokenization. As a result, Okapi BM25 should be interpreted only as a coarse lexical indicator rather than a calibrated similarity measure.

Table~\ref{tab:combined-side-by-side} shows that Agentic PRs have slightly higher central tendency across all four metrics. The differences are modest, but they suggest that agent-generated descriptions align at least as well, and in some cases slightly better, with the content of their edits compared to human descriptions.

\begin{keytakeaway}
\small
\faHandORight \hspace{0.02in} \textit{\textbf{Key takeaway:} Agentic PR descriptions exhibit description-to-diff consistency that is comparable to, and in some metrics slightly higher than, Human PRs, with the clearest differences appearing in the semantic similarity measures.}
\vspace{-7pt}
\end{keytakeaway}

\section{Implications}
RQ1 shows that differences between Agentic and Human PRs arise not only in the amount of code changed, but in how changes are structured and distributed. Review processes and triage heuristics may therefore benefit from using commit count and files touched as early indicators of an Agentic PR's scope, rather than relying solely on LOC-based measures. Deletion-heavy or wide-scope Agentic PRs may warrant closer review, although further work is needed to link these structural patterns to concrete risks. Variability across agent types further suggests that Agentic PR behavior is not uniform, and that modeling agent identity and task context may be important when studying downstream outcomes such as review effort, latency, or defect likelihood.

For RQ2, the consistently high semantic alignment between descriptions and diffs suggests that agent-authored PR description is generally coherent with the code edits it describes. This level of consistency may support downstream tasks that depend on accurate summaries, such as release-note generation, change-log construction, or reviewer routing. Automated checks could further highlight cases where description–diff alignment is unusually low, enabling developers to intervene when an Agentic PR is potentially ambiguous or misleading.

\section{Related Work}
Research on AI-assisted programming has examined models such as Codex and GitHub Copilot, focusing on code completion, developer productivity, and interaction patterns~\cite{vaithilingam2022expectation,vaithilingam2023copilot,barke2023grounded,ogenrwot2025patchtrackcomprehensiveanalysischatgpts,ogenrwot:2024ase}. Most of this work investigates snippet-level behavior rather than project-scale contributions such as pull requests.

Software engineering studies on human-authored patches have analyzed patch size, structure, and review effort~\cite{gousios2014pull,10.1145/2884781.2884840,wang2025automated,watanabe2025useagenticcodingempirical,ogenrwot:2025scam} and their relationships to maintainability and defect risk~\cite{rahman2013predicting,10992485}. Empirical analyses of AI-generated patches remain limited, though early work reports refactoring~\cite{horikawa2025agentic}, stylistic repetition, boilerplate generation, and occasional inconsistencies between descriptions and code changes~\cite{Tambon,horikawa2025agentic,wu2025empirical}.
The AIDev dataset~\cite{li2025aidev} enables the first large-scale, PR-level analysis of how AI agents author and describe code changes in real-world repositories, addressing a gap in existing literature.

\section{Threats to Validity}

\textbf{Internal validity.} Our results depend on the completeness of the AIDev dataset and the reconstructed Human-PR commit data. Missing or truncated patches may bias the distributions, although PRs lacking patch text were excluded. GitHub API retrieval for Human PRs may introduce gaps due to rate limits, deleted repositories, or rewritten histories. Additionally, it may truncate diffs for very large changes, which can reduce patch completeness and may affect similarity metrics computed from diff text.

\textbf{Construct validity.} Our structural metrics (additions, deletions, commits, files touched) approximate PR scope but do not capture intent or correctness. Lexical and semantic similarity measure description–diff alignment, but not code quality or reviewer understanding. Okapi BM25 is unbounded and sensitive to tokenization, so it should be interpreted only as a relative lexical indicator.

\textbf{External validity.} Findings generalize to open-source GitHub projects represented in the AIDev dataset and to the agent versions included in it. Results may differ in private or industrial settings, with different review cultures, or as coding agents evolve.

\section{Conclusion}
We conducted a large-scale empirical comparison of Agentic and Human PRs in the AIDev dataset. Agentic PRs differ most clearly in commit structure and the breadth of modified files, while differences in added or changed lines are relatively small. For RQ2, both PR types show high semantic alignment between descriptions and code, with Agentic PRs exhibiting slightly stronger consistency. These findings suggest that AI-generated PRs are structurally distinct yet generally coherent in how they describe their edits. Future work should examine how these patterns relate to review effort, defect risk, and maintainability as coding agents and development workflows continue to evolve.
\vspace{-12pt}
\begin{acks}
This research was supported by the National Science Foundation Grant Award No.~2519136. We also acknowledge National Science Foundation MRI Grant (\#2117941) for GPU Cluster support.
\end{acks}

%
\bibliographystyle{ACM-Reference-Format}
\bibliography{references}


\end{document}